\documentclass[useAMS,usenatbib]{mn2e} 
\usepackage{natbib}
\usepackage{graphicx}
\usepackage{rotating}

\usepackage{txfonts}

\def\Msun{M$_{\odot}$}

\title{The NGC 4013 tale: a pseudo-bulged, late-type spiral shaped by a major merger}

\author[Jianling Wang et al.]{ Jianling Wang$^{1,2}$\thanks{E-mail:wjianl@bao.ac.cn},
                               Francois Hammer$^{3}$\thanks{E-mail:francois.hammer@obspm.fr},
                               Mathieu Puech$^{3}$,
                               Yanbin Yang$^{3}$,
			       Hector Flores$^{3}$ \\
$^{1}$Key Laboratory of Optical Astronomy, National Astronomical Observatories, Chinese Academy of Sciences \\
$^{2}$National Astronomical Observatories, Chinese Academy of Sciences, Beijing 100012, China.\\
$^{3}$Laboratoire GEPI, Observatoire de Paris, CNRS-UMR8111, Univ. Paris-Diderot, 5 place Jules Janssen, 92195 Meudon France. }

\begin{document}

\date{Received ; accepted }


\maketitle
 
\begin{abstract}
  Many spiral galaxy haloes show stellar streams with various
  morphologies when observed with deep images. The origin of these
  tidal features is discussed, either coming from a satellite infall
  or caused by residuals of an ancient, gas-rich major merger. By
  modeling the formation of the peculiar features observed in the NGC
  4013 halo, we investigate their origin. By using GADGET-2 with
  implemented gas cooling, star formation, and feedback, we have
  modeled the overall NGC 4013 galaxy and its associated halo
  features. A gas-rich major merger occuring 
   2.7 to 4.6 Gyr ago succeeds in reproducing the NGC 4013
  galaxy properties, including all the faint stellar features, strong gas warp,
  boxy-shaped halo and vertical $3.6{\mu}m$ luminosity distribution.
  High gas fractions in the progenitors are sufficient to reproduce the
  observed thin and thick disks, with a small bulge fraction, as
  observed. A major merger is able to reproduce the overall NGC 4013
  system, including the warp strength, the red color and the high
  stellar mass density of the loop, while a minor merger model can 
  not. Because the gas-rich model suffices to create a pseudo-bulge
  with a small fraction of the light, NGC 4013 is perhaps the
  archetype of a late-type galaxy formed by a relatively recent
  merger. Then late type, pseudo-bulge spirals are not mandatorily
  made through secular evolution, and the NGC 4013 properties also
  illustrate that strong warps in isolated galaxies may well occur at
  a late phase of a gas-rich major merger.
\end{abstract}


\section{Introduction}

Mergers are important for galaxies formation and evolution in the concordance
cosmological model. Semi-empirical simulations and observations have shown that
about half local spiral progenitors have experienced a major merger process
(mass ratio greater than 1:4) since $z \sim 1.5$
\citep{hammer2005,hammer2009, puech2012}.  {The process of the disk
rebuilding after a major merger has been extensively explored by modern
state-of-art simulation
\citep{robertson2006,hopkins2009,sh2005,hammer2010,wang2012}.
\citet{hopkins2010} using large set of simulations showed that the final disk
fraction of major merger remnant is a function of mass ratio and gas fraction, although
it shows a large scatter due to the merger orbital parameters (see their Fig. 7). For example, equal (or 3:1) mass merger can
 lead to a new spiral with bulge fraction lower than 28\% (12\%) as long as the
gas fraction is higher than 80\% (50\%), respectively. Since high redshift galaxies have much higher gas
fraction than local spirals \citep{rodrigues2012}, this is consistent with the
disk rebuilding scenario after a major merger \citep{hammer2005}.  } The
imprints of major merger events are easily and rapidly washed out in the center
part of galaxies, because the density is high and the dynamical timescales are
short. But at large scales, in the outskirts of galaxies, the density is low,
and the dynamic timescales are expected to be relatively large, which can let
imprints of the merger process lasting several Gyr. Then ancient major merger
events should leave imprints in spiral halos, if many spirals have seen their
disk rebuilt after a gas-rich major merger.

Deep observations by \citet{martinez2010} have revealed that many
local spirals show faint tidal features in their halo (see e.g., M31
and NGC 5907) with diverse morphologies. These faint features have
been formerly interpreted as evidence for minor mergers. 
However, an issue with this scenario is that the nucleus of the
  infalling galaxy is difficult to destruct. Such a residue is
  generally not observed, even though it is predicted by simulations
\citep{martinez2008, martinez2009}: this is a clear weakness for the
minor merger alternative.
 
\citet{wang2012} have shown that the complex two loops in the
halo of NGC 5907 can be well reproduced by a major merger model.
\citet{hammer2010} also modeled the M31 Giant Stream as being formed
by tidal tail particles captured by the remnant of an ancient major
merger. More modeling is needed to show that gas-rich major mergers 
can generate the various tidal features observed in local spiral
halos, as it is predicted that about half local spirals have
experienced such events \citep{hammer2009,puech2012}.

In this paper, we use a gas rich major merger model for reproducing
the loop and the other faint features observed in the halo of NGC 4013 as
well as the overall structure of this edge-on spiral galaxy. Deep
observations obtained by \citet{martinez2009} have revealed many faint
features including a loop-like feature in the stellar halo of NGC
4013, which has been formerly interpreted as caused by a minor merger
event despite the absence of any progenitor residue. Deep
observations with Spitzer IRAC by \citet[hereafter C11]{comeron2011}
have revealed that there is an extra flattened extended component in
the vertical direction lying besides the thin and the thick disks. Such
a component contains $\sim 20\%$ of the total mass and C11
convincingly argued that it cannot be explained by an ongoing minor
merger. Furthermore the HI map of NGC 4013 reveals a very prominent
warp. The above motivated us to verify whether a major merger can  
reproduce all the peculiarities observed in NGC 4013, including the
low bulge fraction, the loop system, the warp, and the extended
vertical component.

This paper is organized as follows. In Section 2, we describe the
basic properties of NGC 4013, and summarize the observed
peculiar features. The simulation method and initial conditions are
described in Section 3. The results are presented in Section 4. In
the last Section, we discuss our results and summarize them.

\section{NGC 4013 properties}
\label{sec:property}
\subsection{The NGC 4013 galaxy}

NGC 4013 is an Sb type edge-on spiral \citep{buta2007}, and there are
thirteen redshift-independent distances available for this
galaxy, ranging from 15.6 to 24.1 Mpc, with an average value of 18.6
Mpc. Using the baryonic Tully-Fisher relation, this results in a
velocity $\sim$ 210 km/s, while the observed rotational velocity
of NGC 4013 is about 177 km/s \citep{mcgaugh2005}, with a maximum
velocity of 195 km/s \citep{bottema1996}. Therefore, we adopted a
distance of 16.9 Mpc \citep{willick1997}, which is consistent with the
baryonic Tully-Fisher relation \citep{mcgaugh2005,puech2010} and
within one sigma from the average distance calculated above.

There is a boxy bulge component that is thought to be a bar or
the result of a merging event \citep{am2002,bp1985}. There are
different estimates of the bulge fraction in the literature.
\citet{bianchi2007} used a radiative transfer model to decompose the
$V$ and $K'$ band images into a stellar disk, bulge, and dusty disk to
account for extinction. We considered only B/T determination in NIR
bands to avoid the strong extinction associated to an edge-on galaxy.
\citet{bianchi2007} found that the bulge fraction is 27\% in K band by
fixing the S{\'e}rsic index of the bulge component to 4.
\citet{mcdonald2009}, using one-dimension fitting in $K$ band, found a
value of only 1\%. The S4G
project\footnote{http://www.oulu.fi/astronomy/S4G\_PIPELINE4/MAIN/index.html\#entry1003}\citep{sheth2010}
has released the result of two components fitting on Spitzer IRAC
3.6 ${\mu}$m images with a S{\'e}rsic bulge and an exponential disk. In
their result, the bulge fraction is about 9\%. We fitted the Spitzer 
IRAC 3.6 ${\mu}$m image with different components using GALFIT 3
\citep{peng2010}. For comparison with C11, we used similar components
except we added one bulge profile. In our fitting, one S{\'e}rsic
bulge, two edge-on disks, and one exponential component for the stellar halo
are used. We found that the bulge, thin disk, thick disk, and stellar halo
fractions are 4\%, 25\%, 53\%, and 19\%, respectively.

The thick disk is about a factor 2.12 larger than
  the thin disk. This fraction agrees with C11, who found that the
thick disk is $1.2-2.2$ larger than the thin disk, and
that the extended component contains 20\% to 26\% of the total mass
without considering the bulge component. We found that the S{\'e}rsic
index of the bulge is 0.12, indicating that the bulge of NGC
4013 is a pseudo bulge. The disk scalelength of our fitting is 2.11 kpc for both the thin
and thick disks, which is comparable to \citet{saha2009} who got
a value 2.3 kpc by fitting the Spitzer IRAC 4.5 $\mu$m image. We
notice that the optical scalelength is larger than the infrared one,
which is found to be 3.24 kpc \citep{Kruit1982, martinez2009}.

The total stellar mass was estimated following \citet{wang2012} using  
 M/L-color relation from \citet{bell2003} with the 2MASS $K_s$
band and B-V color from NED after correcting for Galactic extinction.
A $''$diet$''$ Salpeter IMF was assumed. Then we get that the stellar
mass of NGC 4013 is about $4.4 \times 10^{10}$ M$_{\odot}$. The HI
mass is about $2.8\times$ 10$^9$ M$_{\odot}$ from \citet{bottema1996}
and the H2 mass is about 1.9$\times 10^9 $M$_{\odot}$ from Gomez de
Castro \& Garcia-Burillo(1997). This leads to a total gas mass of about 6.6
$\times 10^9 $ \Msun $~$ after considering a 40$\%$ He contribution
\citep{just2006}, and a total gas fraction of 13\%.

\subsection{Peculiar features of the NGC 4013 halo}

From its mass and gas content NGC 4013 has been considered as a galaxy
relatively similar to the Milky Way \citep[hereafter MD09]{martinez2009}.
However, observations show that NGC 4013 is very different from the Milky Way
in many aspects:

\begin{itemize}

\item There is a "prodigious" HI warp
  starting at the optical edge of NGC 4013, which is one of the largest
  warps ever observed \citep{bottema1987, bottema1996, bottema1995};

\item A giant loop-like stellar stream was found by MD09, among the
  brightest one ever observed \citep{martinez2008};
 
\item A box-shaped outer "stellar halo" (larger than 15 kpc) was also
  found surrounding the galaxy (MD09), together with four faint tidal
  features;

\item C11 identified a massive extended halo component, which contains
  about 20\%-26\% of the total mass. The mass of the thick disk is about 1.2
  to 2.2 times larger than the thin disk, contrasting with the Milky Way,
  for which this ratio is only 0.21 \citep{read2008}.

\end{itemize}

\subsection{Brightness and red color of the NGC 4013 loop}

The observed mass surface density of the loop can provide useful
constraints on the model. MD09 measured a B-R color of
$1.6_{-0.4}^{+0.6}$, and ${\mu_R} = 27.0_{-0.2}^{+0.3}$. With these
numbers one can derived the mass surface density for the loop
structure, which is ${\Sigma}_{\star}=1.42_{-0.93}^{+4.13}$
M$_{\odot}$ ${\rm pc^{-2}}$.

The B-R color of the main loop and associated tidal features is 1.5
after accounting for Galactic extinction and k-correction. It
corresponds to an old metal-rich stellar population,
intermediate between that of elliptical (B-R= 1.62) and lenticular
(B-R= 1.43) galaxies (see Table 3 of \citealt{fukugita1995}).
Comparing to observed galaxies, it is indeed redder than what is found
for field S0 galaxies after accounting for Galactic extinction and
k-correction \citep{barway2005}.

Let us consider that the loop can be either related to a massive
progenitor or alternatively to a red dwarf elliptical. It is $\sim$
0.6 mag redder than the Sagittarius stream
\citep{martinez2004}. This challenges the stream origin as an old,
metal poor stellar population related to dwarf ellipticals. From its
derived stellar surface density, the NGC4013 loop is 1.6 mag brighter
than the NGC5907 loops and 4 mag brighter than the Sagittarius stream.
The above strengthens our motivation to model it as resulting from a
prominent tidal tail caused by an ancient major merger, and linked to
old and metal-rich stars commonly found in massive galaxies. At the
same time, we aim to also investigate whether such a major merger is
able to explain the strong warp, the boxy halo, and the extended
component.

\section{Simulations and initial conditions}
\label{sec:simu}

{We use the parallel TreeSPH code GADGET-2 \citep{springel2005}, in
which both energy and entropy are conserved.  Cooling, star formation, and
feedback processes of the interstellar medium (ISM) were  implemented as in
\cite{wang2012}, which follow the prescription described by \citet{cox2006} and
\citet{springel2000}. Here we give brief introduction of the method (see \citealt{wang2012} for more details). Radiative cooling
is important for gas to cool down to the central region of the dark halo and
then form stars. The cooling rate is calculated following \citet{katz1996}, in
which the gas is treated as a primordial plasma and the ionization states of H
and He and collisional ionisation equilibrium are assumed.  During the
simulation, gas particles in these dense regions can form stars,  and the star
formation rate of each gas particle is calculated according to
Kennicutt-Schmidt law ($\frac{{\rm d}\rho_{\star}}{{\rm
d}t} \propto \rho_{gas}^{1.5}$, $\rho$ standing for local gas density, see \citealt{kennicutt1998}).   As stars form, gas is transformed into
collision-less matter using a stochastic technique \citep{springel2003}.
Supernova feedback is associated with star formation and is crucial for the
stability of the simulations and regulating star formation.  The energy from
supernovae is deposited into the ISM. This energy heats and pressurizes the gas,
and stabilizes it from further gravitational collapse. The feedback energy can
also be thermalized, which is controlled by the time scale of this 
process and the equation of state.}

The initial conditions were set up similarly to \citet{wang2012},
except that we scaled down the scalelength of the two merger
progenitors by a factor two to account for the size evolution with
redshift \citep{franx2008,wuyts2010}. This scaling factor is
consistent with a later-type galaxy size evolution relation 
\citep[][see Table 1]{Wel2014}.
Progenitors are 60\% $-$ 70\% gas-rich spirals, which is reasonable
for z=1.5-2 galaxies (corresponding to 9 Gyr ago, see
\citealt{rodrigues2012}). We used the same initial baryonic fraction
($f_{bar}$= 9\%) and progenitor density profiles that what was adopted
by \citet{wang2012}.

Cosmological simulations provide constraint on the orbital
parameters. The pericenter and eccentricity are chosen to be
following \citet{kb2006}.
Briefly, we estimated the virial radius of the primary galaxy halo 
at the time of merger occuring ($\sim$ 7 Gyr ago) to be about 114 kpc, 
 following the relationship between the virial radius and total mass derived 
from cosmological simulations \citep{springel1999}. The pericenter 
 used in our
simulations was larger than $\sim$ 23 kpc. Then the ratio between
pericenter and virial radius is 0.20 and the eccentricity is larger
than 0.86, which satisfies the cosmological constraints \citep{kb2006}.

We choose for the primary galaxy a retrograde and closed to polar
orbit, and for the secondary galaxy a prograde and inclined orbit. A
retrograde orbit is favorable for an efficient disk rebuilding 
\cite[see e.g.,][]{hammer2010}, and well-suited to NGC 4013, which is a
spiral with small bulge component. The prograde orientation of
  the secondary is favorable for generating long tidal tails
\citep{wang2012,hopkins2009,tt1972,bh1992}.

{Given the large parameter space, we acknowledge not having been able to thoroughly examine all possible configurations.
However, thanks to the above constraints, the searching parameter space can be
 significantly reduced. In our former work \citep{wang2012}, the theory of  
the tidal tail particles motion in galaxy potential have been extensively explored,
which help optimize the searching parameter space. During the searching
process, we first use low resolution simulation with a few hundred thousand
particles to primarily find the possible geometry. This primary search was done
in a large coarse grid of parameters consisting of pericenter distance,
eccentricity of orbit, and inclination angles for these two interacting
galaxies. After getting the possible parameters, we finely tune the parameters
around this point with one or two million particles. We confirmed that the
results between low and high resolution simulations show little differences,
except the later leads to a better definition of the loop and tidal tail features.  The softening length adopted in low resolution
for dark matter, gas, star particles are, 0.3 kpc, 0.15 kpc, and 0.1 kpc.  In
general, increasing particles number $N$ allows the gravitational softening
length to be decreased with increasing computational time \citep{cox2006}. We
have tested the softening effect by decreasing softening length for some high
resolution simulations as shown in Table 1, and the results do not differ too
much. This indicates that our simulations are converging well. 
We found that a series of models in a family of a parameters with
pericenter between 23 kpc and 27 kpc, and eccentricity of orbit between 0.9 and
0.93, can produce the overall NGC4013 (see Table 1). There is a degeneracy in
the models of NGC 4013 similar to that NGC 5907 \citep{wang2012}, which is the
loop order can not be identified based on current observations. Further deep 
observations are need to definitively break this degeneracy (see the following section). 
}


\section{Results}
\label{sec:result}
\subsection{Loop formation during the merger}

\citet{wang2012}  described the basic properties of loop systems
formed after a major merger, which have lead to successfully model NGC
5907 and of its two loops structures. \cite{hammer2013} 
demonstrated that the dSphs in the M31 disk of satellites are
spatially distributed along a similar system of loops. In this paper,
we use the loop properties to model the tidal features observed in NGC
4013.

\begin{figure*}
\centering
\includegraphics[width=18cm]{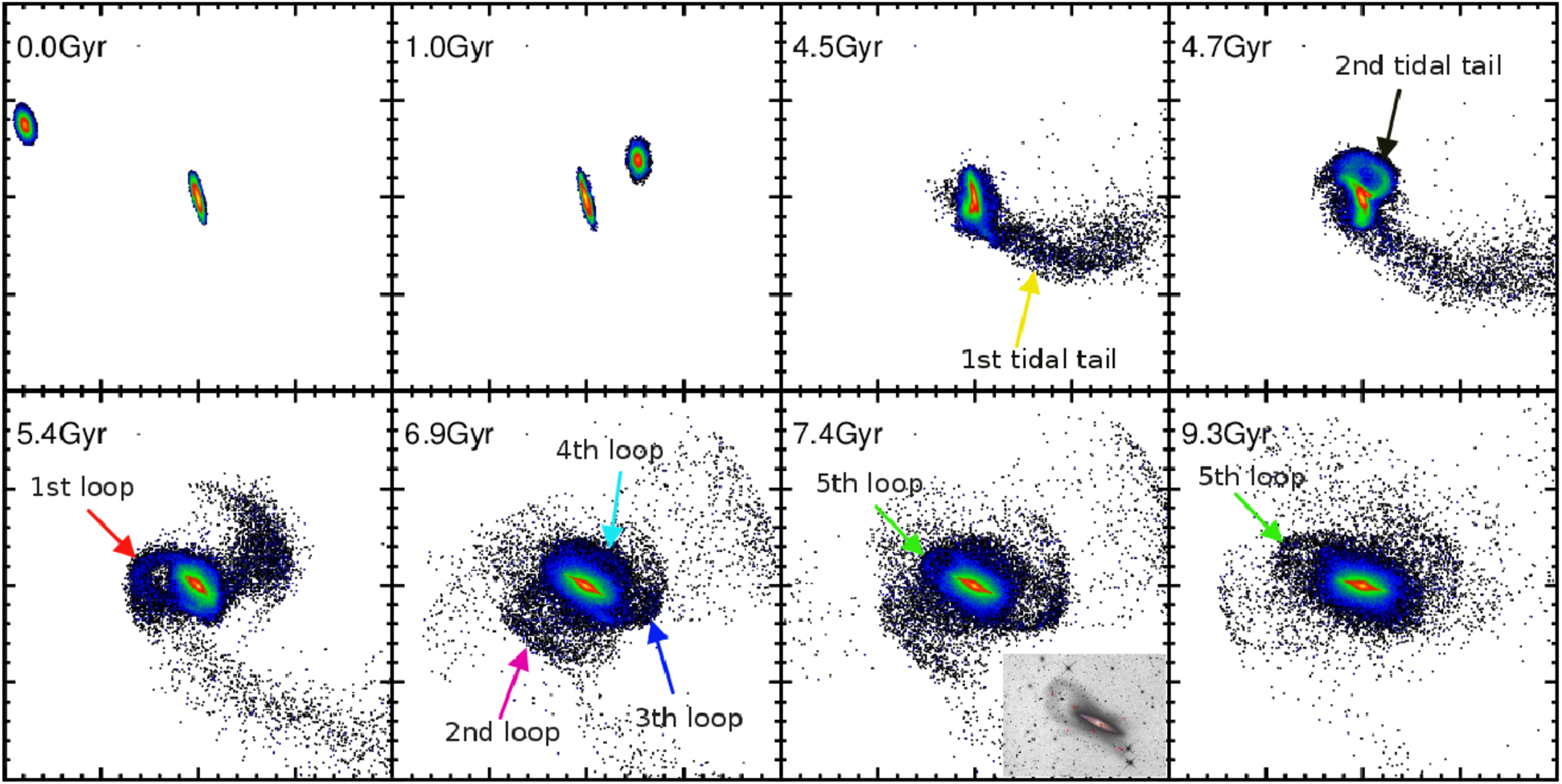}
\includegraphics[width=18cm]{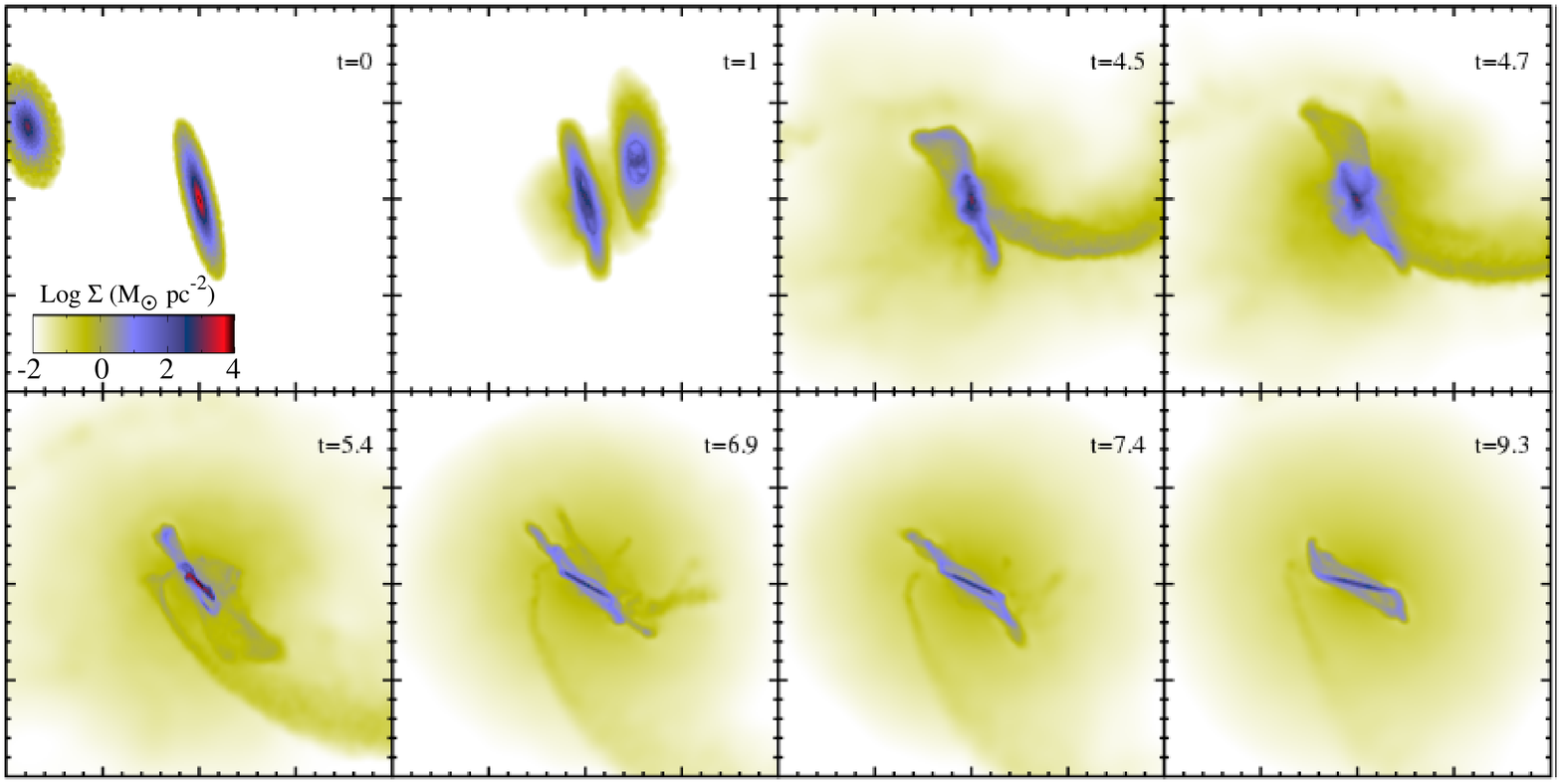}
\caption{Star and gas particles distributions for a merger with mass
  ratio 1:3 (model M3R24M2) at different epochs. Star particles are
  shown in the top two rows. The best fit to observations is at 7.4
  Gyr, at which the structure and over-density of the fifth loop also
  match the observations, {and the observed image from \citet{martinez2009} is shown 
  at the right bottom of this panel.} The gas surface density is shown in the
  bottom two rows. The size of each panel is 200 by 200 kpc and the
  arrows are explained in the text.}
\label{fig:image}
\end{figure*}

\begin{figure*}
\centering
\includegraphics[width=18cm]{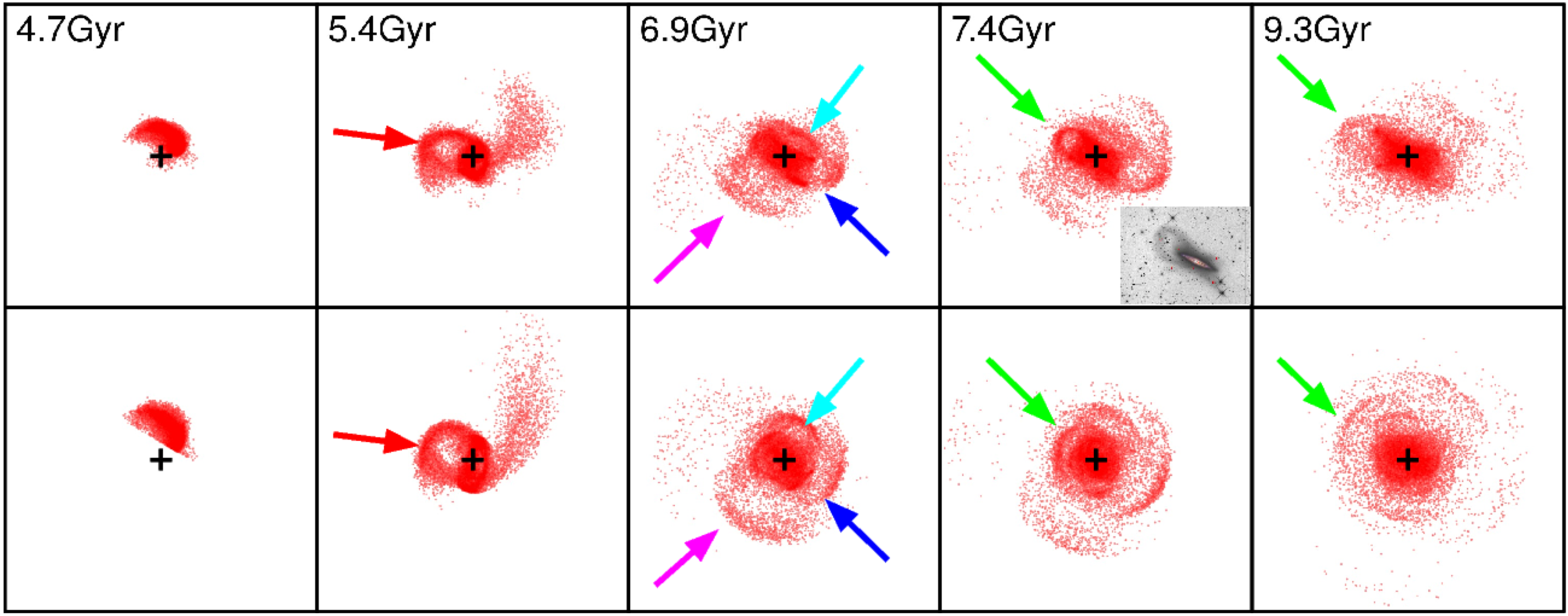}
\caption{Loops formed by the second tidal tail for model M3R24M2. Only
  particles in the tidal tail are shown here. The top row shows the
  loops formation process at the viewing angle matching NGC 4013. The
  loop at 7.4 Gyr resembles that of NGC 4013, while according to its
  brightness the loop at 9.3 Gyr is only marginally consistent with
  the observed one. The bottom row shows the same data with a face-on
  view. At 5.4 Gyr, the red arrow shows the first loop formed. The
  second, third, and fourth loop are formed at 6.9 Gyr indicated by
  purple, blue, and light blue arrows, respectively. The fifth loop,
  which resembles observation at 7.4 Gyr is labeled by a green
  arrow as well as in the 9.3 Gyr panel. {Observed image is overlaped at the 7.4 Gyr panel.}}
\label{fig:loop}
\end{figure*}

The top two rows of Figure \ref{fig:image} show the stellar
distribution of a 1:3 major merger. In the 4.5 Gyr panel, the first
tidal tail is generated after the first passage, as indicated by a
yellow arrow. The second tail is formed before fusion at 4.7 Gyr and
is indicated by a black arrow. This tidal tail will generate the loop
observed in NGC 4013. The first loop forms at 5.4 Gyr and is shown by
a red arrow. It is followed by the second, third, and fourth loops,
which are shown in the 6.9 Gyr panel and are labeled by a purple, a
blue, and a light blue arrow, respectively. At 7.4 Gyr the fifth loop
(labeled by a green arrow) is formed and resemble what we observed
in NGC 4013. As shown by \citet{wang2012} the loops become larger hence
fainter with time. The loop system can be maintained during at least 9.3 Gyr. To
better illustrate how the loops formed through potential captured from tidal tail
particles, we isolated them and traced their motions in Fig. \ref{fig:loop}.

The loop formation mechanism is similar to that described in
\citet{wang2012}. New loops are continuously fed by falling particles with time, 
 while low order loop become fainter, which make them more difficult to be
detected than high order loops.

Since the loops appear consecutively, from low order to high order
with different surface brightnesses (or mass) densities, it helps identify
the one matching the observations. Meanwhile, tuning the initial position 
angle of the primary galaxy control the final disk position relative 
to a given order loop. The phase angle of the different loops are close 
to 90 degrees \citep{wang2012}, so generally we just needed to rotate the 
disk by 180 degrees to have the loop and disk geometry matching observation 
for a given loop order. For details about the loop formation and properties, 
refer to \citet{wang2012}.

\begin{figure*}
\centering
\includegraphics[width=3.1cm]{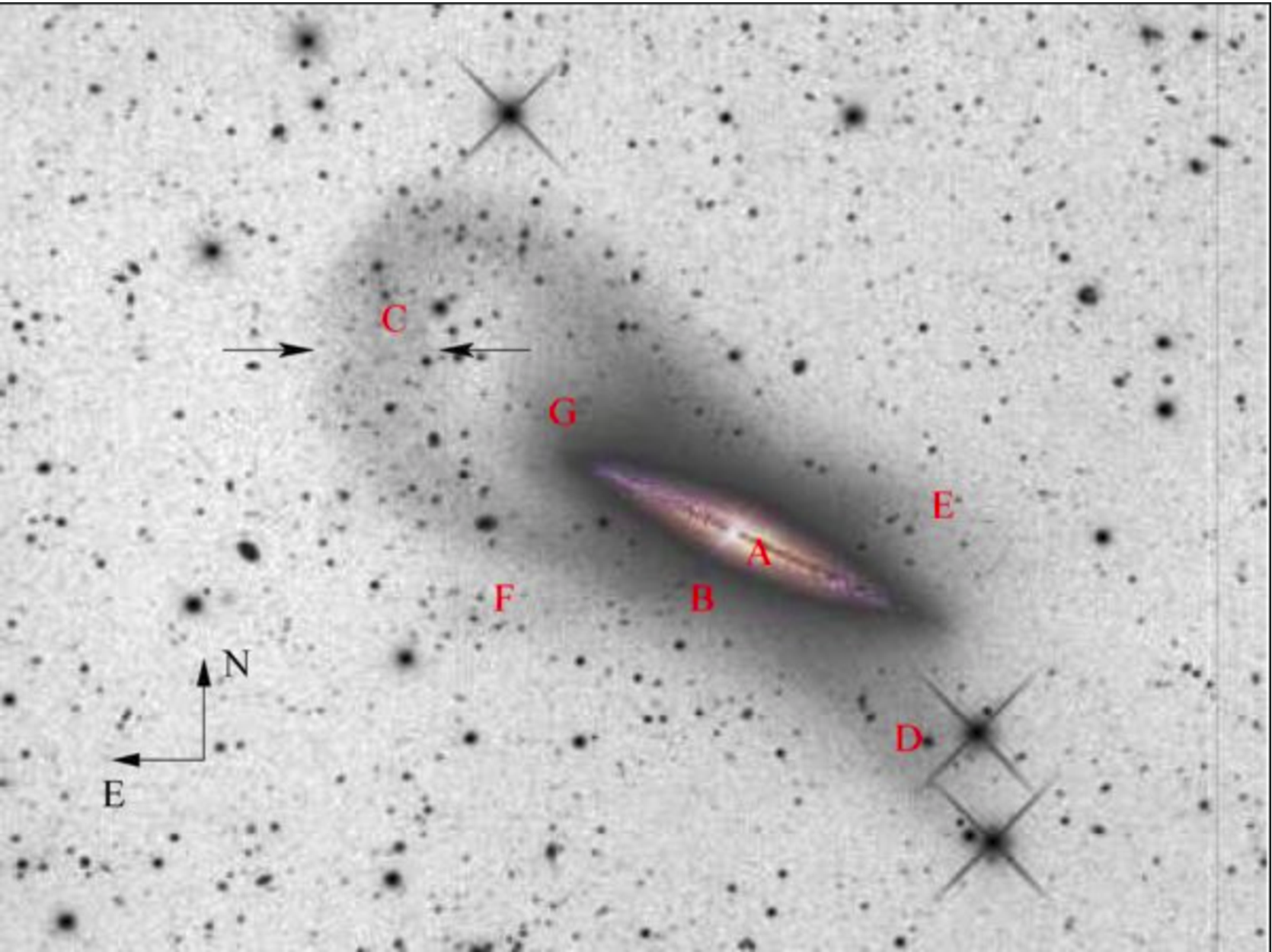}
\includegraphics[width=13.2cm]{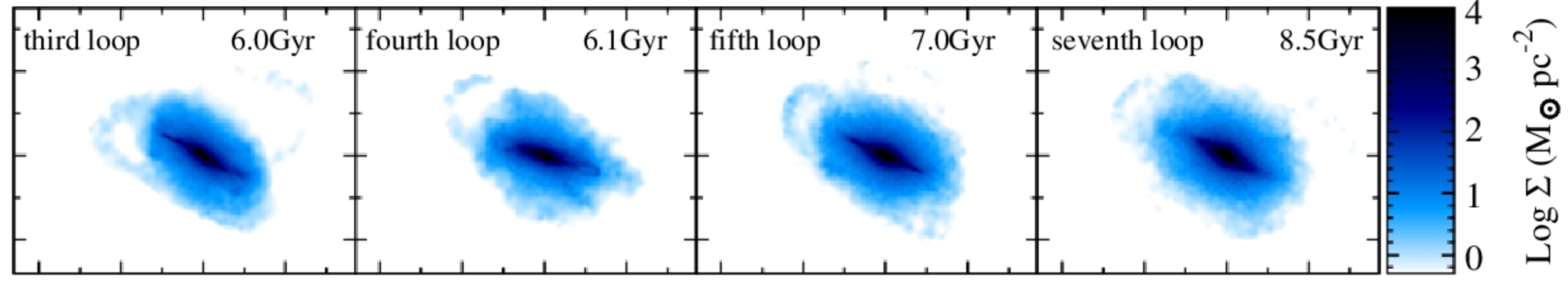}
\caption{Comparison between the observation predictions of different order
loops formed at different time. {The left panel is observed image \citep{martinez2009}, while 
the right four panels are from simulation.} The surface mass density is cut at
${\Sigma}_{\star}= 0.5 ($M$_{\odot}~{\rm pc^{-2}})$, which corresponds to the
$1\sigma$ limit of loop surface mass density. The high order loop appears much
later compared to those of low order.  The size of each panel is $\sim $ 82 by
62 kpc. Models used from left to right are M3R23M2, M3R23M1G7, M3R26M2A, and
M3R27M2. }
\label{fig:looporder}
\end{figure*}

In Fig. \ref{fig:looporder}, we show different loop orders formed by particles 
falling back onto the remnant, which show geometrical resemblance with the
observations. Four different models, M3R23M2, M3R23M1G7, M3R26M2A, and M3R27M2 
are used to produce these different loop orders at the different times. The loop 
order and the observed time are labeled at top of each panel. 
 Each panel shows a mass surface density that is limited
by the observed loop density, i.e., ${\Sigma}_{\star}= 0.5$ M$_{\odot} 
{\rm pc^{-2}}$.  From current observations, it is difficult to judge what is the 
loop order that best fits the observations because we need more constraints
on their formation time. This is similar to that was found by \citet{wang2012}.

The two bottom rows of Figure \ref{fig:image} show the 
  distribbution of the gaseous phase during the merger process. At
6.9 Gyr there is formation of a strong warp induced by the merger, which becomes
stronger with time. At 7.4 Gyr and 9.3 Gyr it persists and resembles
the observed one. The warp of NGC 4013 is one of the strongest
observed one \citep{bottema1987}, which indeed suggests a not too
ancient, violent merger event.

\subsection{Box-shaped feature formed by particles falling back from the same tidal tail}

MD09 (see their Figure 3) identified a huge, at least 3\arcmin
($\sim$15kpc) sized box-shaped in the outer stellar halo, which they
called feature B. The loop-like feature (C for MD09) starts at the
northeast corner of the boxed-shaped halo, and ends at the southeast
corner of the box. There is also a pair of wings, one at the northwest
direction (E) and a more extended one (D) is at the southwest. Here we
show that the box-shaped stellar halo and the four wings out of the
disk are a natural product of single major merger model, and are all
coming from the same system of tidal tails.

\begin{figure*}
\includegraphics[width=4.2cm]{martinez_fig3.ps}
\includegraphics[width=12.4cm]{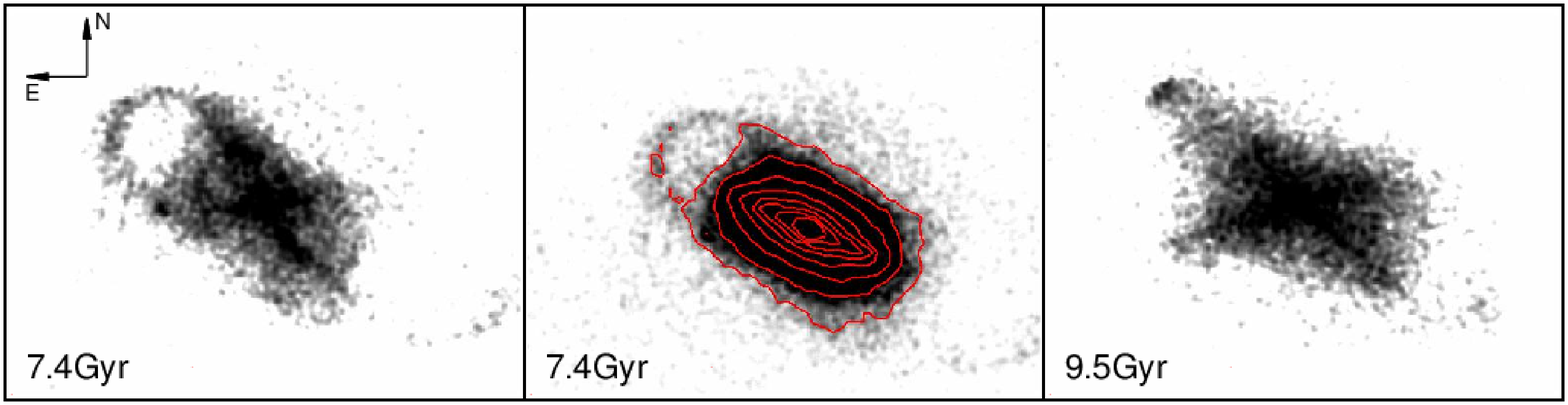}
\caption{{The left panel show observed image from \citet{martinez2009}, and the right three panels show results from simulation.} Box-shaped feature formed by the coming back particles in the
  tidal tail of model M3R24M2. Here the contrast is tuned to clearly
  view the X-shaped morphology. 
  {In the second and right panels, the }
  particles are selected by isolating tidal tail particles after the
  formation of the second tidal tail, and tracing their motions to 7.4
  Gyr and 9.5 Gyr. The X-shaped features is not a transient feature
  but would require deep color image to be detected. It generates the
  box-shaped stellar halo and a pair of (observed) wings features west
  and southwest. In the third panel all particles are shown with
  contours overlapped to evidence the box-shaped feature in the
  stellar halo resulting from the tidal tail particles. The size of
  each panel is $\sim $ 82 by 62 kpc. }
\label{fig:xshape}
\end{figure*}

Fig.\ref{fig:loop} shows stellar particles from the tidal tails only,
at different epochs. After their capture, stellar particles move
  along loops that are mostly lying in a single plane, though it
thickness can be affected by precession. 
{The second and fourth panels }
of Fig.\ref{fig:xshape}, we show their projected surface mass density,
at 7.4 Gyr and 9.5 Gyr. It reveals a clear X-shaped surface density
that is predicted in our model because tidal tail particles are coming
from 4 directions drawn by the two almost symmetric loops seen in the
7.4 Gyr panel of both Figures \ref{fig:image} and
\ref{fig:loop}.
The X-shaped feature persists for several Gyr as loops and is similar
to what was found by \cite{purcell2010}. 
The huge box-shaped stellar halo in NGC 4013 results from the
accumulation of particles surrounding the X-shaped feature (see
contours in Fig.\ref{fig:xshape}, {third} panel).
Then our modeling naturally explains the box-shaped halo (B), the loop
(C) and the pair of wings lying west of the disk (D and E
features for MD09).

As explained and detailed in \citet{wang2012}, the high order loops
are made of material coming originally from the inner region of the
progenitors compared to low order one, they are associated with larger stellar mass surface
densities. This property suffices to explain the relatively high
surface brightness of the boxy halo as well as that of the loop if
those correspond to high order loop particles, i.e., with a relatively
recent merger. In the next sub-sections we verify whether
this is consistent with other observations of NGC4013.

\subsection{Strong gas warp generated by the major merger process}

\begin{figure*}
\centering
\includegraphics[width=14cm]{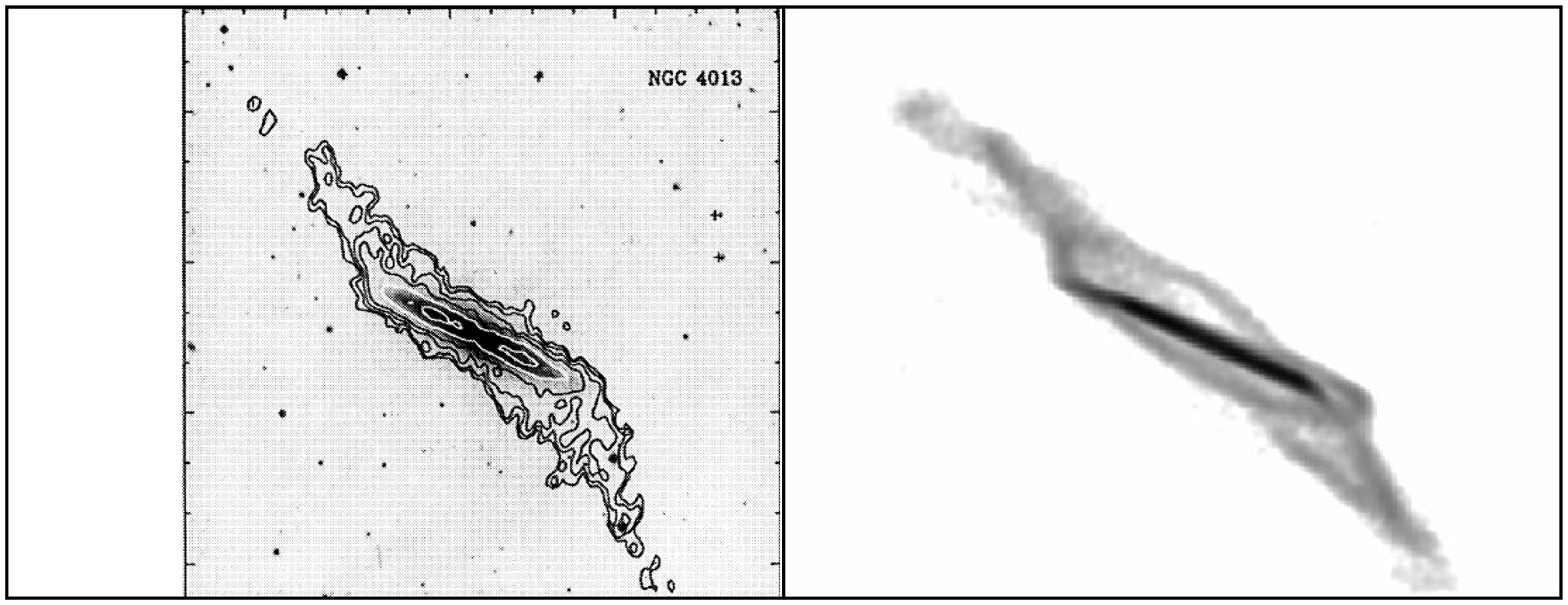}
\caption{Comparison between the observed gas morphology with that from
  the simulation. In the left panel the contours show the observed HI
  map superimposed on the optical image (from \citealt{bottema1996}),
  and the right panel shows the simulated mass density of cold gas
  from model M3R24M2.
  The size of each panel is $\sim $ 82 by 62 kpc. }
\label{fig:compgas}
\end{figure*}

Due to the dissipational properties of the gas, the gas motions in the halo are
rapidly decoupled from stellar streams (see Figure \ref{fig:image}). Most major
merger events result in gas warps, which can last for several Gyr as shown in
Fig.\ref{fig:image}. The angular inclination of the massive progenitor has been
set close to polar, and with a retrograde orbit this results in a very strong
gas warp. Fig. \ref{fig:compgas} shows a comparison between the observed HI gas
morphology from \citet{bottema1996} with the simulation for which only cold gas
($T < 20,000 K$) is shown.

Observations show that the gas fraction of NGC 4013 is about 13\%.  After
turning the feedback strength down after fusion \citep{wang2012, hammer2010} to
allow a new disk to reform after the major merger, most of the gas is consumed
and converted into stars. This results in a gas fraction between 14\% and 16\%
for most models, which matches well the observed range (see Tab. 1).

\subsection{The NGC4103 disk rebuilt after the major merger}

\begin{figure}
\centering
\includegraphics[width=9cm]{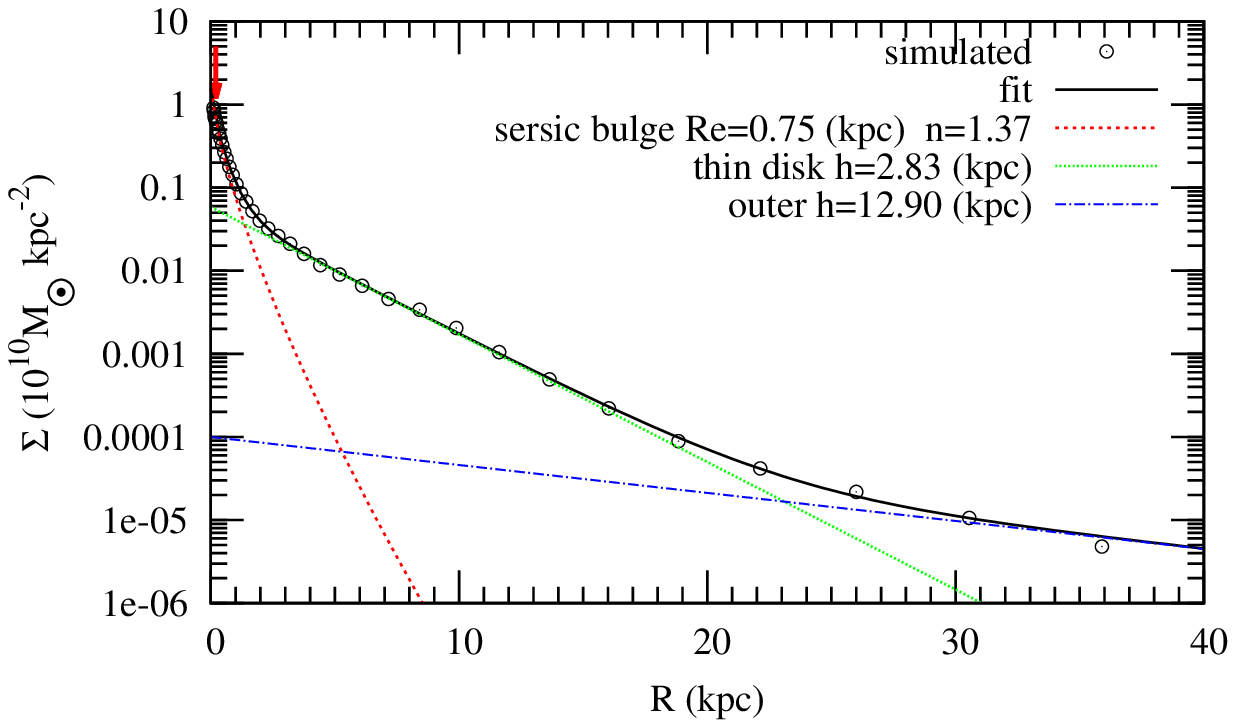}
\caption{Surface brightness mass density distribution as a function of
  radius for model M3R24M2. In the fitting S\'ersic bulge and
  exponential disk components are used. Black open circles indicate
  the simulation results and the black line shows the fitting result.
  Different components are shown using different color lines. The
  red arrow (just above $\Sigma$=1 and R$<<$ 1 kpc) indicates two
  times values of the softening length, which is very small when
  compared to the bulge characteristic size. Then the softening 
  does not affect the result.}
\label{fig:surf}
\end{figure}

As shown in Section 2, NGC 4013 has a pseudo-bulge \citep{Kormendy2004} with a
very small B/T value. To check whether a newly rebuilt spiral galaxy is
consistent with NGC4013, we analyzed the mass surface density in the simulation
as it was done in \citet{wang2012}. A S\'ersic bulge and two exponential disk
components were fitted, while the central region, which is dominated by the
softening was excluded. Fig.\ref{fig:surf} shows the fit for model M3R24M2, and
the fitting results are listed in Table 1. In this model a small bulge fraction
is obtained with B/T $\sim$ 25\% and with a S\'ersic index of 1.37. By
increasing the initial gas fraction to 70\%, see model M3M1G7, the B/T ratio
decreases to 14\%. All the models listed in Table 1 show bulge S\'ersic indices
smaller than 1.7, which are consistent with pseudo-bulges \citep{Kormendy2004}.
The disk scale-length in our model fitting is about 2.8 kpc, which is close to
that observed.

\subsection{Rotation curve}

\begin{figure}
\centering
\includegraphics[width=9cm]{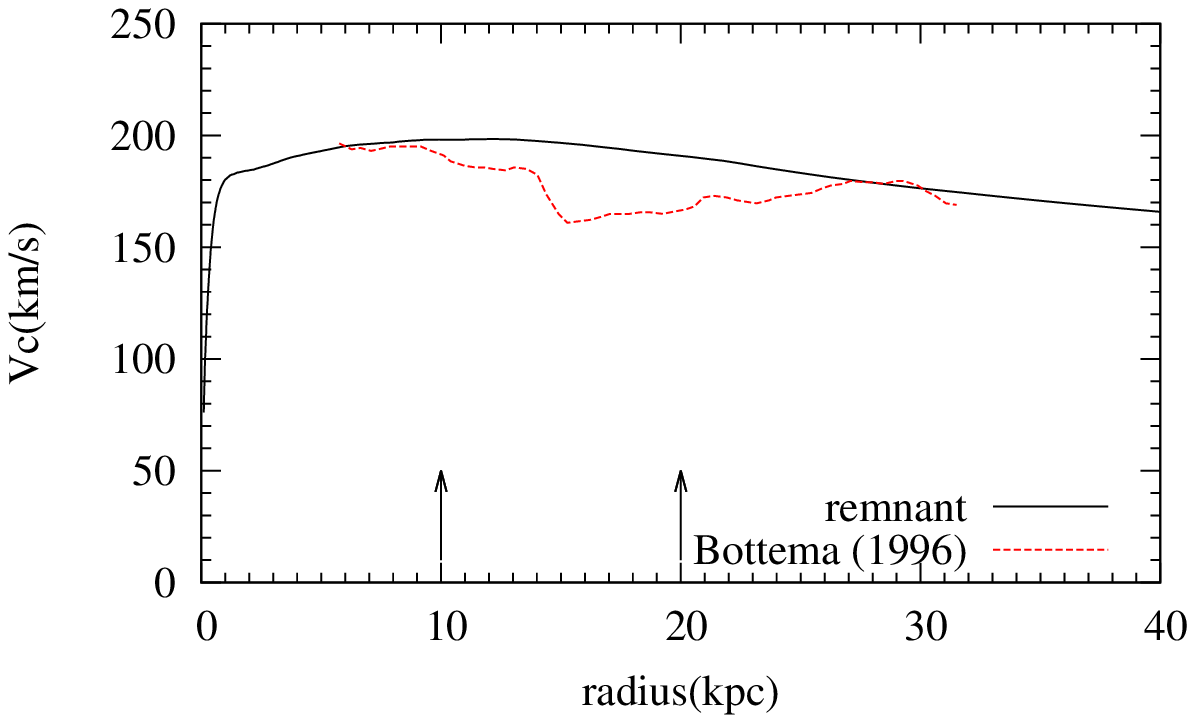}
\caption{Comparison between the rotation curves of the simulated
  galaxy after the major merger (black line, model M3R24M2 at 7.4 Gyr)
  and observation (red dashed line). The observation rotation curve is
  from \citet{bottema1996}. The rotation curve of the simulation is
  calculated by simply using the mass model, with
  V$_c=\sqrt{\frac{GM(<r)}{r}}$. }
\label{fig:rc}
\end{figure}

We checked whether the rotation curve predicted by the model 
matches the observed one. In Fig. \ref{fig:rc}, the observational
data are taken from \citet{bottema1996}, while the model rotation curve is
calculated assuming a rotating disk at equilibrium
(V$_c=\sqrt{\frac{GM(<r)}{r}}$, \citealt{wang2012}). At r $<10$ kpc,
the rotation curve of the model agrees well with observations. From 10
to 20 kpc (as indicating by two arrows), the observed rotation curve 
falls down dramatically, which is due to impact of the strong warp.

\begin{table*}
\begin{center}
\caption{Parameters of the six models used in this study. Feedback model:  high 
 feedback as in \citet{cox2006} is used before fusion, and changed to 
low feedback after fusion. Bulge S{\'e}rsic index $n$ and disk scale-length are obtained by fitting the surface 
brightness density distribution with a S{\'e}rsic profile and two exponential functions.}
{\scriptsize
\begin{tabular}{lccccc}
\hline \hline
parameters                     & M3R23M2& M3R24M2&M3R26M2A & M3R27M2  & M3R23M1G7  \\
\hline                                                                 
mass ratio                     & 3      & 3      &  3      &  3       &  3        \\
Gal1 incy                      & -100   &-105    & -100    &  -100    &  -100      \\
Gal1 incz                      & -130   &-130    & -130    &  -130    &  -130      \\
Gal2 incy                      & -70    &-70     & -70     &  -70     &  -70       \\
Gal2 incz                      & 30     & 30     & 30      &  30      &  30        \\
Gal1 gas fraction              & 0.6    & 0.6    & 0.6     &  0.6     &  0.7       \\
Gal2 gas fraction              & 0.6    & 0.6    & 0.4     &  0.4     &  0.7       \\
$r_{peri}$ (kpc)               & 23     & 24     &  26     &  27      &  23        \\
eccentricity                   & 0.93   & 0.93   &  0.9    &  0.9     &  0.93      \\
N$_{particle}$                 & 2.0M   & 2.0M   &  2.0M   &  2M      &  1M        \\
m$_{dm}$:m$_{star}$:m$_{gas}$  &4:1:2   & 4:1:2  & 4:1:2   &  4:1:2   &  4:1:2     \\
softening($\epsilon_{dm}$:$\epsilon_{star}$:$\epsilon_{gas}$)(kpc)& 0.3:0.1:0.15& 0.3:0.1:0.15&0.24:0.08:0.12&0.24:0.08:0.12 &0.3:0.1:0.15 \\
\hline                                                                          
Observed time (Gyr)            & 7.2    & 7.4    & 7.0     &  8.5     &  6.1       \\ 
Thin disk scalelength(kpc)     & 2.83   & 2.83   & 2.72    &  2.81    &  2.56     \\
Bulge sersic index($n$)        & 1.64   & 1.37   & 1.32    &  1.10    &  0.81      \\
Re of bulge (kpc)              & 0.88   & 0.75   & 0.83    &  0.91    &  0.78      \\
B/T 			       & 23\%   & 25\%   & 21\%    &  21\%    &  14\%      \\
Final gas fraction	       & 14\%   & 14\%   & 16\%    &  16\%    &  28\%      \\
\hline                               
\end{tabular}                        
}
\end{center}                         
\label{tab:par}                      
\end{table*}

\subsection{Vertical Profile}

\begin{figure*}
\centering
\includegraphics[width=13cm,height=7.7cm]{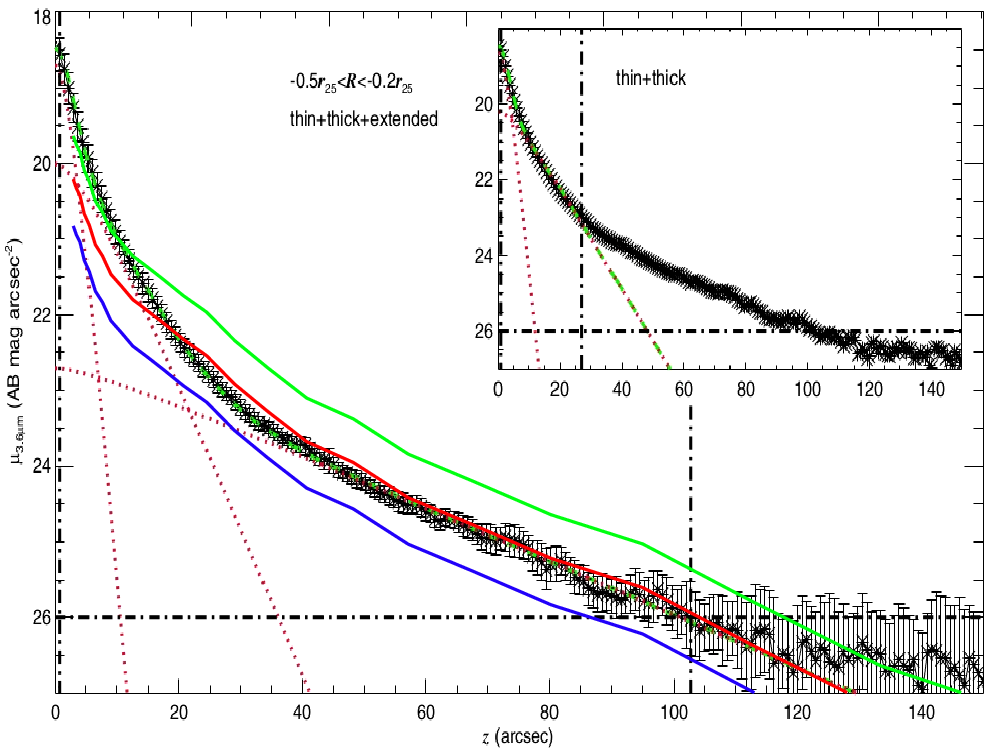}
\caption{Comparison between the observed vertical $3.6{\mu}m$
  luminosity profile with simulations (model M3R24M2). The solid color
  lines from simulations are overlapped on Figure 3 of
  \citet{comeron2011}. The red solid line shows results from
  simulations by assuming mass to light ratio 0.85 \citep{meidt14}. The
  green and blue solid lines show profiles with different mass to
  light ratio of 0.5 and 1.5, respectively while the M/L=0.85
  reproduces well the extended halo component, it might need M/L=0.5
  for fitting both disks. }
\label{fig:zprof}
\end{figure*}

By fitting a vertical luminosity profile on the Spitzer IRAC
$3.6{\mu}m$ image, C11 found that NGC 4013 can not be
fitted satisfactorily by a canonical thin+thick disk structure.
Instead, they found evidence that there are a thin and a thick disk
and an extra flattened component, which contains $\sim 20\%$ of the
total mass. Fig.\ref{fig:zprof} compares our simulation model with the
observed vertical profile from \citet{comeron2011}. To measure the
vertical mass surface density, we have selected the same physical
radius chosen by \citet{comeron2011}, namely $-0.5 r_{25} < R< -0.2
r_{25}$, with $r_{25}=147\arcsec$. To convert stellar mass into $3.6
{\mu}m$ light, we used a stellar mass to light ratio M/L of 0.85 from
\citet{meidt14} by assuming a 'diet' Salpeter initial mass function
\citep{eskew12}. To account for uncertainties, we adopted a range
  of 0.5-1.5 on M/L.

Fig.\ref{fig:zprof} shows the result. The simulation (red curve)
matches  the observations all the way from z=20 to 120$\arcsec$
predicting well the presence of the extra component. The red curve
 corresponding to a mass-to-light ratio of 0.85 at the inner region under-predicts 
the observations, which might be related to a change in M/L between
the different structures (see e.g., the green curve that matches the
observations near the inner region).

\section{Discussion and Conclusion}
\label{sec:discuss}

We modeled NGC 4013 as resulting from a 3:1, gas-rich major
merger. The NGC4013 loop and other tidal features can be explained by
stellar streams, which are natural and persistent residuals of such
mergers. The model suggests that tidal tails stars were  
captured by the gravitational potential of the remnant, and formed a loop
system that have lifetimes maximized by the absence of dynamical
 friction. The models also predict a newly formed disk and a small bulge, which
show many similarities with the observed ones. It also reproduces well
the distribution of stars at the outskirts, including the box-shaped
halo and the light profile along and perpendicular to the disk. The
merger event predicts a strong gas (HI) warp, which is spatially
well decoupled from the old stars, as observed. With our
modeling the above observed features can be reproduced all together if the
merger occurred 2.7 to 4.6 Gyr ago, with a first passage
occurring about 3.5 Gyr earlier.

\begin{figure*}
\includegraphics[width=18cm]{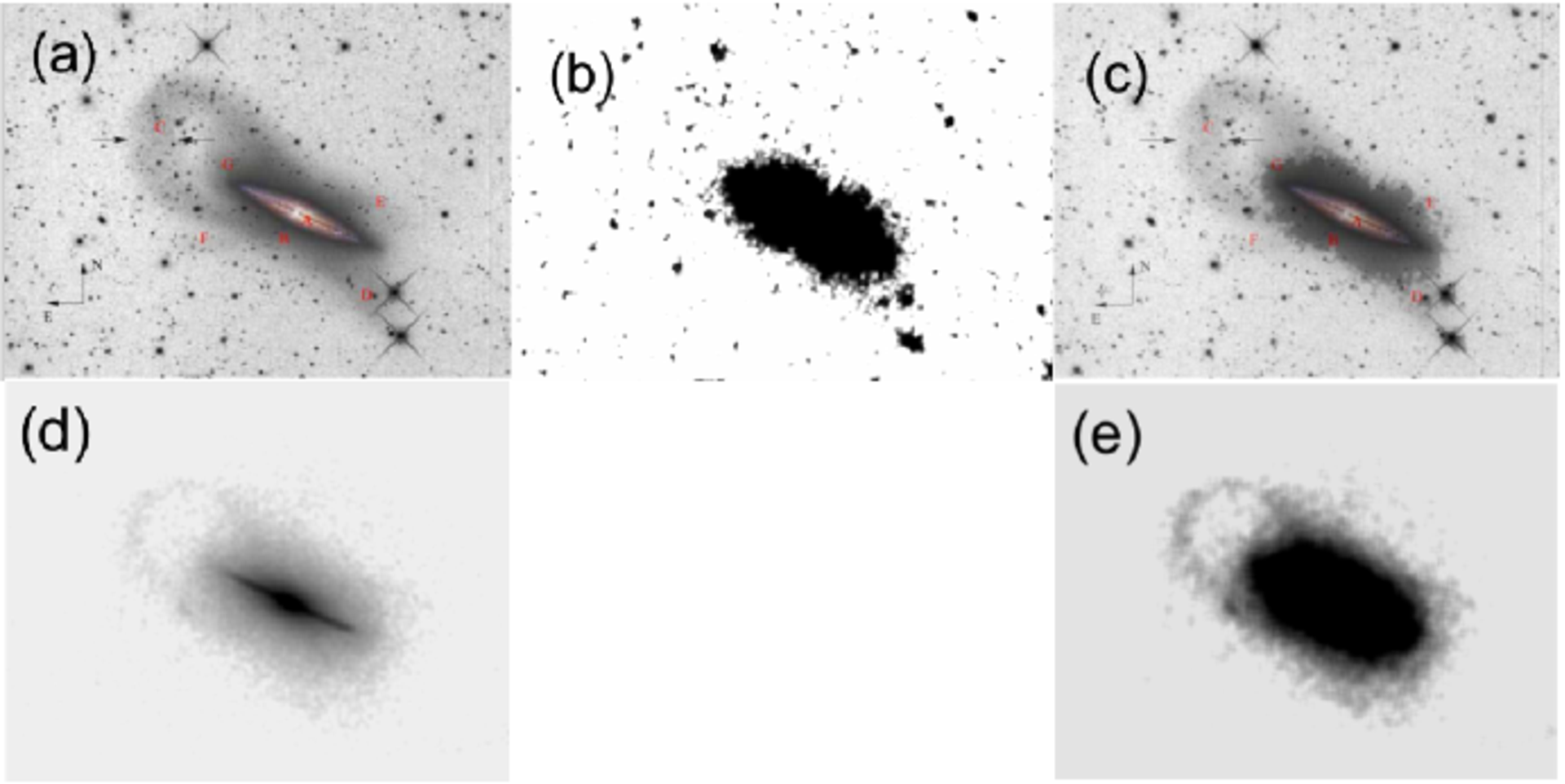}
\caption{Comparison between observations and simulations. Panel (a)  
  shows deep observations at optical wavelengths from MD09,
  revealing clearly the loops and other faint features. Panel (b) 
  shows observations with Spitzer IRAC 3.6${\mu}$m
  evidencing the thick disk and the stellar halo while the loop and
  other features are too faint to appear. Panel (c) shows combination 
  of (a) and (b), which trace the stellar mass distribution less 
  affected by dust extinction effects.  Panel (d) shows the simulation of model M3R24R2 
  with very high contrast showing mostly the bulge, the thin disk and at fainter levels, 
  the loop and the extended halo. Panel (e) shows the same model as (d) with a 
  cut at a lower isophotal level showing the overall halo. The mass surface 
  density of the simulation has been cut at the lower limit of the
  observed loop. The size of each panel is $\sim $ 82 by 62 kpc. 
    }
\label{fig:compobs}
\end{figure*}

{The deepest image available for NGC 4013 is a combination of four
images within different filters. This combined image results from
various stellar populations and is also affected by dust extinction, which hampers a proper comparison with
simulations, for which the stellar mass is the
most reliable quantity. As the near infrared imagery traces better the stellar mass,
we prefer to compare our simulation with near infrared data.  In Fig.
\ref{fig:compobs} we show the deep optical observations, the IRAC 3.6 ${\mu}$m
image, and the simulation, together (compare panels (a) and (c), with panels (d) and (e), respectively).} The mass surface density of the simulation
was cut at the surface brightness limit of the observations.  The deep IRAC
observations show the presence of a thick disk and of an extended stellar halo.
It does not show the loop simply because it is not deep enough to recover it.

An important role of major mergers in the past history of spiral
galaxies is expected from observations of their progenitors, 6
  Gyr ago \citep{hammer2005,hammer2009}, and empirically based 
  $\Lambda$-CDM calculations of the merger rate
\citep{hopkins2010,puech2012}. If galactic disks can be rebuilt after
gas-rich mergers, it is quite natural that they may produce strong
warps, tidal loop systems of old stars, and extended halo components.
 This paper presents the first ever made simulation reproducing
  successfully the overall properties of NGC4013. Previously, MD09
 had superimposed a model aimed at reproducing the Monoceros Stream on
NGC4013 revealing geometrical similarities. MD09 however noticed that
 a more massive progenitor was required to reproduce the
NGC 4013 loop system.

Perhaps the most difficult challenge for a minor merger scenario is
the elliptical-type color (${\rm B}-{\rm R} = 1.6^{+0.6}_{-0.4}$) of
the NGC 4013 loop: it is $\sim$ 0.6 mag redder than metal poor, old
stars that populate dwarfs or dwarf streams. For example dSphs show B
- R $\le$ 1, even for those dominated by old-stars (e.g., UMi,
\citealt{martinez2001} and Sculptor, \citealt{hurley1999}). The NGC
4013 loops and halo are perhaps an extreme illustration of previous
findings \citep{mouhcine2006,zibetti2004} that stellar haloes of most
spiral galaxies (but the MW, see \citealt{hammer2007}) are too red to
be inhabited by metal-poor stars related to dwarfs.

It could be argued that a minor merger is still plausible, if some
systematics affect the color measurement of a low surface
brightness structure such as the NGC 4013 loop (but see MD09). If
correct, the minor merger scenario has also to account for the much
larger stellar mass surface density than what is found in tidal tails
of dwarfs. Moreover a dwarf infall could explain neither the strong
warp, the boxy-shaped halo or the vertically extended component.
For those one would have to invoke independent origins, for example
cooling gas for the warp (e.g., \citealt{radburn2014}) and a 10:1
merger for the halo (e.g., C11 and see Figure 1 of
\citealt{purcell2010}). Adding this to the fact that the supposed
dwarf residual is not seen in the deepest images as could be
expected (see the discussion in \citealt{wang2012}), a merger of
satellite appears very unlikely for interpreting the NGC 4013 system.

\cite{wang2012} tested the range of more massive minor (or
intermediate) merger (from 15:1 to 10:1) for reproducing the NGC 5907
system, using characteristic times similar to NGC 4013. They
estimated that 12:1 is a limit for reproducing NGC 5907 with 
orbital parameters as expected from cosmological simulations, and this
agrees with predictions from \cite{hopkins2009} for the formation of
massive spirals. However, if a $\sim$ 10:1 merger may explain the
box-shaped halo \citep{purcell2010}, it could not account for the red
color of the NGC 4013 loop.

This leaves a major merger as the only possible alternative. NGC 4013 also
appears to show quite exceptional properties when it is compared to
other edge-on spiral galaxies. In C11, it is one of the two (among 46)
galaxies showing an extra vertical component to the thin and thick
disks. It is also the only one (among seven) showing an extended
vertical dust distribution above the disk, which has been detected by
$Herschel$ \citep{Verstappen2013}. We suggest that these exceptional
properties are due to a recent occurrence of the last major merger. In
fact, most of them are expected to occur 6.5 to 8 Gyr ago in massive
spirals (see e.g., \citealt{hammer2009,puech2012}), instead of less
than 4.6 Gyr for NGC 4013. At later time one may expect fainter  
loops, less intense warp, and less exceptional IR properties as the
remnant is progressively relaxing. The loop redness may also support 
a quite recent event, because it would let enough time for
the progenitors to be metal-enriched. In fact, there is still an
important evolution of metal abundances during the past 6 Gyr
\citep{rodrigues2008}.

 Interestingly, we find that NGC 4013, M31 and NGC 5907 would
  follow an order of increasing elapsed time since the last merger
  occurrence, i.e., 2.7-4.6, 5.5 and $\ge$ 6 Gyr, respectively.
\cite{kormendy2010b} have investigated the R=8 Mpc volume and have
shown that 11 among 19 galaxies possess a pseudo-bulge. They argued
that such structures evidence a secular evolution for most spiral
galaxies. At distances of 16.9 and 14 Mpc, respectively, NGC 4013 and
NGC 5907 are not in the volume studied by \cite{kormendy2010b}, but
their properties should be similar. Both show pseudo-bulges that can be  
reproduced by gas-rich major mergers. As discussed in
\cite{hammer2012} while a classical bulge has to be formed through a
violent event such as a major merger, a pseudo bulge does not permit
to do such a distinction. In their realization of several orbits and
feedback histories, they do find that most gas-rich merger remnants
possess pseudo-bulges, a property confirmed by independent simulations
\citep{keselman2012}.

Only a major merger can reproduce all the very particular
properties of NGC 4013, which places this galaxy as a potential
missing link between well relaxed spiral galaxies and $\sim$ 0.5-1 Gyr
post-merger galaxies { such as J033245.11-274724.0 at z=0.43, which dusty forming disk makes it an excellent candidate for an early rebuilding disk phase \citep{hammer2009b}}. Further investigations would be helpful to
verify the above. In fact, our model helps to predict many features
(see Figure \ref{fig:image}) that could be tested in the near future:
\begin{itemize}
\item Deeper observations should reveal many other loops at fainter
  surface-brightness levels; if obtained in two filters, color may also
  helps to identified the X-shaped feature in the halo { as well as more accurate photometry would help to better constrain the age/metal of the stellar population};
\item Deeper HI observations would provide an incomparable test
  between mergers of different mass ratios; 
\item The warp duration time (less than 2 Gyr, see bottom panels of
  Figure \ref{fig:image}) could be verified by estimating the ages 
  of the stars as done by \cite{radburn2014} for the NGC 4565
  HI disk warp;
\item On average the disk stars are predicted to be of intermediate
  ages, i.e., formed 2.7 to 4.6 Gyr after the merger, which could
  be identified through the signatures of the Balmer absorption lines and the
  4000A break\footnote{We recovered a spectrum of NGC 4013 from the SDSS, 
  showing a quite prominent $H\delta$ absorption line, though it only covers 
  the central part of the galaxy}.
\end{itemize} 

As a concluding remark, we notice that NGC 4013 may provide a crucial
test about the long-standing issue of the origin of strong warps in
isolated galaxies. The observed spatial decoupling between the warp
and the tidal stellar stream is supporting a major merger origin, and
this alternative should be considered by future studies.

\section*{Acknowledgments} 
We thank the referee for very useful suggestions that help to improve the paper.  This
work has been supported by the ''Laboratoire International Associ\'e'' Origins,
the Young Researcher Grant of National Astronomical Observatories, Chinese
Academy of Sciences. Computations were done using the special supercomputer at
the Center of Information and Computing at National Astronomical Observatories,
Chinese Academy of Sciences. The images in gas surface density of
Fig.\ref{fig:image} and Fig.\ref{fig:looporder} were produced using SPLASH
\citep{price2007}. 

\label{lastpage}

\end{document}